\documentclass[superscriptaddress,aps,twocolumn,longbibliography]{revtex4-1}
\usepackage{amsmath,amssymb}
\usepackage{graphicx}
\usepackage{dcolumn}
\usepackage[bb=boondox]{mathalfa}
\usepackage{lipsum}
\usepackage{enumitem}
\usepackage{float}
\usepackage{bm,dsfont}
\usepackage{multirow}
\usepackage{simpler-wick}
\usepackage{cuted}
\usepackage{color}
\usepackage{hyperref}
\usepackage{lipsum}
\hypersetup{
colorlinks = true,
linkcolor = [rgb]{0.70,0.13,0.13},
citecolor = [rgb]{0.13,0.55,0.13},
urlcolor = [rgb]{0.25, 0.41, 0.88}}
\newcommand{\bra}[1]{{\langle #1|}}
\newcommand{\ket}[1]{{|#1 \rangle}}
\newcommand{\Var}{\text{Var}}

\newcommand{\ii}{\mathrm{i}}
\newcommand{\id}{\mathbb{1}}

\newcommand{\dsE}{\mathbb{E}}

\newcommand{\scM}{\mathcal{M}}

\newcommand{\scO}{\mathcal{O}}

\newcommand{\Tr}{\operatorname{Tr}}

\newcommand{\vect}[1]{{\bm{#1}}}

\newtheorem{theorem}{Theorem}

\newcommand{\dia}[3]{\raisebox{#3pt}{\includegraphics[height=#2pt]{dia_#1}}}
\newcommand{\eq}[1]{\begin{equation}#1\end{equation}}
\newcommand{\eqs}[1]{\begin{equation}\begin{split}#1\end{split}\end{equation}}
\newcommand{\eqnref}[1]{Eq.\,\eqref{#1}}
\newcommand{\figref}[1]{Fig.\,\ref{#1}}

\newcommand{\appref}[1]{Appendix\,\ref{#1}}
\newcommand{\refcite}[1]{Ref.\,\cite{#1}}

\usepackage[mathlines]{lineno}

\begin{document}


\title{Hamiltonian-Driven Shadow Tomography of Quantum States}
\author{Hong-Ye Hu}
\affiliation{Department of Physics, University of California San Diego, La Jolla, CA 92093, USA}
\author{Yi-Zhuang You}
\affiliation{Department of Physics, University of California San Diego, La Jolla, CA 92093, USA}

\date{\today}

\begin{abstract}
Classical shadow tomography provides an efficient method for predicting functions of an unknown quantum state from a few measurements of the state. It relies on a unitary channel that efficiently scrambles the quantum information of the state to the measurement basis. Facing the challenge of realizing deep unitary circuits on near-term quantum devices, we explore the scenario in which the unitary channel can be shallow and is generated by a quantum chaotic Hamiltonian via time evolution. We provide an unbiased estimator of the density matrix for all ranges of the evolution time. We analyze the sample complexity of the Hamiltonian-driven shadow tomography. For Pauli observables, we find that it can be more efficient than the unitary-2-design-based shadow tomography in a sequence of intermediate time windows that range from an order-1 scrambling time to a time scale of $D^{1/6}$, given the Hilbert space dimension $D$. In particular, the efficiency of predicting diagonal Pauli observables is improved by a factor of $D$ without sacrificing the efficiency of  predicting off-diagonal Pauli observables. 
\end{abstract}

\pacs{Valid PACS appear here}

\maketitle


\emph{Introduction.}---
Quantum state tomography \cite{Vogel1989Determination, James2001Measurement, Caves2002Unknown, Paris2004Quantum, Roos2004Bell,PhysRevLett.124.100401} is an essential quantum technology underlying the characterization of quantum devices and the discrimination of quantum states. It aims to reconstruct the density matrix from repeated measurements of identically prepared copies of a quantum system. While the complexity of exact tomography of the full density matrix scales exponentially with the system size due to the curse of dimensionality\cite{7956181}, approximate tomography with polynomial complexity has been developed with assumptions of the underlying quantum state, including matrix product state tomography\cite{Cramer2010Efficient,Lanyon2017Efficient,Wang2017Scalable}, reduced density matrix tomography\cite{Linden2002Almost,Linden2002The-Parts,Diosi2004Three-party,Chen2012Comment,Chen2012From,Xin2019Local-measurement-based}, and machine learning tomography\cite{Torlai2018Neural-network,Torlai2018Latent,Carrasquilla2018Reconstructing,Xu2018Neural,Quek2018Adaptive,Torlai2019Integrating,Carrasquilla2019Probabilistic,Cha2020Attention-based,Neugebauer2020Neural-network}. 

\begin{figure}[htbp]
\begin{center}
\includegraphics[width=0.6\columnwidth]{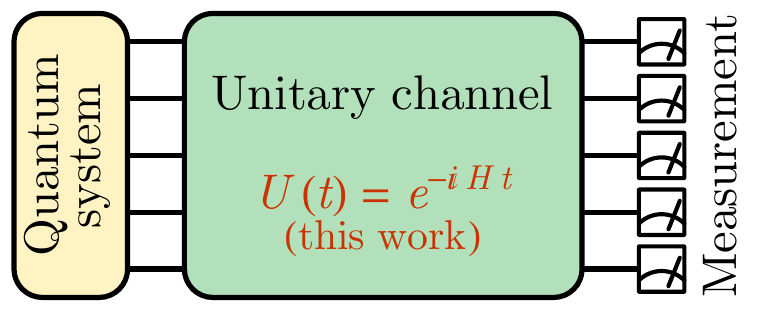}
\caption{Shadow tomography protocol. Specially in this work, the unitary channel is considered to be a time evolution generated by a random Hamiltonian $H$ for time $t$.}
\label{fig:protocol}
\end{center}
\end{figure}

Among various tomography schemes, \emph{shadow tomography}\cite{Brandao2017Quantum,Aaronson2017Shadow,Aaronson2019Gentle,Huang2020Predicting,Zhao2020Fermionic,Koh2020ClassicalSW,Elben2020Mixed-State,2020arXiv201109636C,Zhou2020Single-Copies,Struchalin2021Experimental,Garcia2021Quantum} has recently attracted much research attention. Given a copy of a $N$-qubit quantum system described by the density matrix $\rho$, the shadow tomography protocol first performs a random unitary transformation $U$ on the state $\rho\to\rho'=U\rho U^\dagger$, then measures the transformed state $\rho'$ in the computational basis (i.e.~simultaneously measuring the Pauli-$Z$ operator on every qubit), as illustrated in \figref{fig:protocol}. After the measurement, the system collapses to a pure state $\ket{b}$ labeled by the bit-string $b\in\{0,1\}^N$ of measurement outcomes. Classical snapshots $\hat{\sigma}=U^\dagger \ket{b}\bra{b}U$ of the quantum system can be collected through repeated measurements. Given the knowledge about the random ensemble of the unitary transformation $U$, the density matrix $\rho$ can be reconstructed as a particular linear combination of the ensemble average of classical snapshots $\hat{\sigma}$, where the linear channel only depends on properties of the unitary ensemble.

The unitary transformation $U$ plays an important role in the protocol to scramble the quantum information, such that the computational basis measurement on the scrambled state $\rho'$ is effectively a simultaneous measurement of a random set of $N$ commuting operators $\{U^\dagger Z_i U\}_{i=1}^{N}$ on the original state $\rho$. In this way, each measurement returns measurement outcomes for $2^N$ observables (as products of arbitrary subsets of the commuting operators, e.g.~$U^\dagger (Z_iZ_j\cdots) U$), which provides an efficient way to extract information from the quantum system. The shadow tomography saturates the fundamental lower bound on the minimal number of independent samples required for tomography,\cite{7956181,Huang2020Predicting} achieving the maximal efficiency. In particular, when the unitary ensemble is Haar random (or any unitary 2-design such as random Clifford circuits), the scrambling is strongest to enable estimating rank-1 observables (such as quantum fidelity) with a constant number of samples that does not scale with the system size $N$.

However, insufficient scrambling power of the unitary channel will affect the efficiency of shadow tomography. To investigate the effect of scrambling power on the tomography efficiency, we consider the random unitary channel $U(t)=e^{-\ii H t}$ to be the time evolution generated by a chaotic Hamiltonian $H$, which enables us to tune the scrambling power of the unitary channel by the evolution time $t$. Such a tomography scheme will be called the \emph{Hamiltonian-driven shadow tomography}. When $t=0$, the unitary channel is an identity that has no scrambling power, then the shadow tomography is only able to reconstruct the diagonal part of the density matrix $\rho$ (in the computational basis). As the quantum system evolves for some time, the computational basis measurement will be able to probe the off-diagonal part of $\rho$ more efficiently, but it also becomes less efficient to infer the diagonal part of $\rho$ as the diagonal information starts to get scrambled with the off-diagonal information. In the long-time limit, the unitary ensemble approaches the Haar limit, and the efficiency for both diagonal and off-diagonal parts converges to the same limit. In this work, we derive the reconstruction channel for the Hamiltonian-driven shadow tomography and analyze its efficiency as a function of the evolution time $t$ and the total Hilbert space dimension $D=2^N$. We find that, given observables are Pauli operators, there exists an intermediate time range $1\lesssim t\lesssim D^{1/6}$, in which the Hamiltonian-driven shadow tomography only need $\sim \Tr(O_d^2)/D$ samples to estimate a diagonal observable $O_d$, and $\sim \Tr(O_o^2)$ samples to estimate an off-diagonal observable $O_o$, which is superior to the shadow tomography based on 2-design random unitaries. Our result may find applications in shadow tomography with shallow quantum circuits, which are feasible in the noisy intermediate-scale quantum (NISQ)\cite{Preskill2018Quantum} era.

\emph{Brief Review of Shadow Tomography.}--- Consider a $N$-qubit system described by an unknown density matrix $\rho$ which can be prepared repeatedly. The goal of the quantum state tomography is to infer $\rho$ from repeated measurements on independent copies of the state. In each experiment, the state $\rho$ is first evolved by a unitary operator $U$ drawn from the distribution $P(U)$ and then measured in the computational basis. The measurement will prepare a classical snapshot of the quantum system $\hat{\sigma}\equiv U^\dagger \ket{b}\bra{b}U$ with $b\in\{0,1\}^N$ labeling the measurement outcome. After $M$ repeated experiments, we will collect $M$ classical snapshots $\{\hat{\sigma}_1,\cdots,\hat{\sigma}_M\}$. We can view the average mapping from the quantum state $\rho$ to its classical snapshots as a measure-and-prepare quantum channel $\scM$,
\begin{equation}\label{eq:channel_general}
\scM(\rho)=\dsE\hat{\sigma}=\dsE\left[U^{\dagger}\ket{b}\bra{b}U\right],
\end{equation}
where the averaging is taken over both the unitary ensemble $P(U)$ and the possible measurement outcomes following the probability distribution $P(b|U)=\bra{b}U\rho U^\dagger\ket{b}$. 

The quantum state $\rho$ can be reconstructed by applying the inverse channel $\scM^{-1}$ (i.e.~the reconstruction channel)
\begin{equation}
\rho = \scM^{-1}( \dsE \hat{\sigma})=\dsE \scM^{-1}(\hat{\sigma})=\dsE \hat{\rho},
\end{equation}
where $\hat{\rho}\equiv\scM^{-1}(\hat{\sigma}) = \scM^{-1}(U^\dagger \ket{b}\bra{b} U)$ is called the \emph{classical shadow} of the original quantum state $\rho$. The reconstruction channel $\scM^{-1}$ does not admit physical implementation, as it is not completely positive in general. Nevertheless, given the distribution $P(U)$ of the unitary ensemble, the reconstruction channel $\scM^{-1}$ can be calculated and applied to the classical snapshots $\{\hat{\sigma}_1,\cdots,\hat{\sigma}_M\}$ by classical post-processing\cite{Huang2020Predicting} to obtain the set of classical shadows $\{\hat{\rho}_1,\cdots,\hat{\rho}_M\}$, which can then be used to estimate both linear and nonlinear functions of the underlying quantum state $\rho$. For example, the expectation value $o=\Tr(O\rho)$ of a physical observable $O$ is a linear function of $\rho$, which can be estimated as
\begin{equation}
o=\Tr(O\rho)=\mathbb{E}[\Tr(O \hat{\rho})]\simeq\frac{1}{M}\sum_i \Tr(O \hat{\rho}_i)\label{eq:linear_function}.
\end{equation}
Nonlinear functions, such as $\Tr(O\rho\otimes\rho)$, can also be estimated similarly,
\begin{equation}
\begin{split}
\Tr(O\rho\otimes \rho)&=\Tr(O\mathbb{E}[\hat{\rho}]\otimes \mathbb{E}[\hat{\rho}])\\
&\simeq\frac{1}{M(M-1)}\sum_{i\neq j}\Tr(O\hat{\rho}_i\otimes \hat{\rho}_j)\label{eq:nonlinear_function},
\end{split}
\end{equation}
given the fact that $\hat{\rho}_i$ and $\hat{\rho_j}$ are statistically independent. When number of experiments $M$ is large enough, the statistical averages over classical shadows in \eqnref{eq:linear_function} and \eqnref{eq:nonlinear_function} will converge to their corresponding expectation values without bias.

\emph{Hamiltonian-Driven Shadow Tomography.}--- 
The shadow tomography approach crucially relies on the realization of the unitary channel with sufficient scrambling power. Tomography schemes using global Haar/Clifford unitary ensemble have been proposed in \refcite{Huang2020Predicting}. In practice, it remains challenging to realize these unitary ensembles on NISQ devices. We propose to generate the scrambling unitary channel by some quantum chaotic Hamiltonian $H$ through time-evolution $U(t)=e^{-\ii H t}$. 
With this setup, the quantum dynamics also enter the discussion, as it becomes meaningful to discuss how the reconstruction channel $\scM^{-1}$ and the tomography efficiency depends on the evolution time $t$.

To analyze the problem, we model the chaotic Hamiltonian generally by a random Hermitian operator $H$ drawn from the Gaussian unitary ensemble (GUE) in $D=2^N$ dimensional Hilbert space, following the probability distribution $P(H)\propto \exp(-\frac{D}{2}\Tr H^2)$. The energy scale is such normalized that the spectral density of $H$ approaches the Wigner semicircle law $\rho(E)=\frac{1}{2\pi}\sqrt{4-E^2}$ of the spectral radius $2$ as $D\to \infty$. The unitary time-evolution generated by $H$ admits the following eigen decomposition
\begin{equation}\label{eq:U(t)}
U(t)=e^{-i H t}=V\Lambda (t)V^{\dagger}.
\end{equation}
where $V$ is the matrix diagonalizing $H$ and $\Lambda(t)$ is the diagonal matrix $\Lambda_{nn'}(t)=e^{-i E_n t}\delta_{nn'}$ with $E_n$ being the eigen energies of $H$. For the GUE random matrix $H$, the unitary $V$ is Haar random.

Substitute the eigen decomposition of $U$ in \eqnref{eq:U(t)} to \eqnref{eq:channel_general}, the quantum channel $\scM$ can be expanded as
\eqs{
\mathcal{M}(\rho)=\mathop{\dsE}_{V,\Lambda}\sum_{b\in\{0,1\}^N}&V\Lambda(t)^\dagger V^\dagger\ket{b}\bra{b}V\Lambda(t) V^\dagger\\
&\bra{b}V\Lambda(t) V^\dagger\rho V\Lambda(t)^\dagger V^\dagger\ket{b}.}
Using the results of Haar measure integral\cite{Weingarten1978Asymptotic,Collins2006Integration} and the spectral form factor of GUE matrices\cite{Cotler2017Chaos}, to the leading order of $D$, the quantum channel simplifies to
\eqs{
\mathcal{M}(\rho)&=\mathcal{M}(\id/D+\rho_o+\rho_d)\\ 
 &=\dfrac{\mathbb{1}}{D}+\dfrac{\rho_o}{\alpha_D(t)}+\dfrac{\rho_d}{\beta_D(t)},
 \label{eq:quantum_channel}
}
where $\id$ stands for the identity matrix, $\rho_o$ is the off-diagonal part of $\rho$, and $\rho_d$ is the traceless diagonal part of $\rho$. The coefficients $\alpha_D(t)$, and $\beta_D(t)$ are defined as
\eqs{\label{eq:alphabeta}
\alpha_D(t) &= \Big(\dfrac{1}{D+1}-\lambda_D(t)\Big)^{-1},\\
\beta_D(t)&=\Big(\dfrac{1}{D+1}+D\lambda_D(t)\Big)^{-1},\\
\lambda_D(t)&=\dfrac{(Dr^2(t)+r(2t))^2-4r^{2}(t)}{(D+3)(D^2-1)},}
and $r(t)=J_1(2t)/t$ with $J_1$ being the Bessel function of the first kind (which captures the leading-$D$ behavior of spectral form factors). See \appref{appen:diagrammatic_channel} for detail derivations.

In the short time limit $t\rightarrow 0$, the quantum channel becomes
\begin{equation}
\mathcal{M}(\rho)\rightarrow \dfrac{\mathbb{1}}{D}+\rho_d.
\label{eq:short_time_limit}
\end{equation}
As expected, only the diagonal part of the density matrix will be transmitted through this quantum channel. Because in the absence of time-evolution, the computational basis measurement can only extract the diagonal information of the density matrix, and the off-diagonal information is completely lost. As the measurement is not tomographically complete, the channel $\scM$ is not invertible at $t=0$. However, after a finite time-evolution, a finite fraction of the off-diagonal information will be scrambled to the diagonal part of the density matrix and become accessible to the measurement, then the channel $\scM$ will be invertible (as long as $t\neq 0$). In the long-time limit $t\rightarrow +\infty$, the quantum channel converges to the known result\cite{Guta2018Fast} for shadow tomography with unitary 2-design
\begin{equation}
\mathcal{M}(\rho)\rightarrow \dfrac{\mathbb{1}}{D}+ \dfrac{\rho_d}{D+1}+\dfrac{\rho_o}{D+1}.
\end{equation}
The diagonal and off-diagonal parts converge to the same channel transmission rate in this limit, indicating that the quantum information has been fully scrambled.

Since $\id/D$, $\rho_o$, and $\rho_d$ are all orthogonal to each other, the inverse channel of \eqnref{eq:quantum_channel} is simply obtained by inverting the coefficient of each term,
\begin{equation}
  \hat{\rho}=\mathcal{M}^{-1}(\hat{\sigma}) = \dfrac{\mathbb{1}}{D}+\alpha_D(t) \hat{\sigma}_o + \beta_D(t) \hat{\sigma}_d\label{eq:inverse_channel},
\end{equation}
where $\hat{\sigma}_o$ and $\hat{\sigma}_{d}$ are respectively the off-diagonal and the traceless diagonal part of the classical snapshot $\hat{\sigma}=U(t)^\dagger\ket{b}\bra{b}U(t)$. Coefficients $\alpha_D(t)$ and $\beta_D(t)$ were defined in \eqnref{eq:alphabeta}. Given the reconstruction channel $\scM^{-1}$ for the Hamiltonian-driven shadow tomography, we can use classical shadows $\hat{\rho}=\scM^{-1}(\hat{\sigma})$ to reconstruct the density matrix $\rho=\dsE \hat{\rho}$ and to estimate physical properties of the quantum system following \eqnref{eq:linear_function} and \eqnref{eq:nonlinear_function}.

\emph{Tomography Efficiency Analysis.}---
We now analyze the efficiency of the Hamiltonian-driven shadow tomography, i.e.~how many independent copies of $\rho$ are typically needed to predict functions of $\rho$ to a suitable precision. We will mainly focus on the efficiency of predicting linear functions, but our result can be generalized to nonlinear function predictions systematically.

According to \eqnref{eq:linear_function}, the linear function $o=\Tr(O\rho)$ can be estimated from the classical shadow $\hat{\rho}$ via $o=\dsE[\Tr(O\hat{\rho})]=\dsE\hat{o}$,
where $\hat{o}\equiv\Tr(O\hat{\rho})$ can be viewed as a random variable derived from the classical shadow. In practice, we conduct $M$ experiments to collect classical shadows $\hat{\rho}_i$, and estimate $o$ using
\eq{o_{\text{avg}}=\dfrac{1}{M}\sum_{i=1}^{M} \hat{o}_i=\dfrac{1}{M}\sum_{i=1}^{M}\Tr(O\hat{\rho}_i).}
Based on Chebyshev's inequality, the probability of the estimation $o_{\text{avg}}$ to deviate from its expectation value $o$ is bounded by its variance $\Var(o_{\text{avg}})$ as
$\text{Pr}(|o_{\text{avg}}-o|\geq \delta)\leq \Var(o_{\text{avg}})/\delta^2$.
To control the deviation probability within a desired statistical accuracy $\epsilon$, we require $\Var(o_{\text{avg}})/\delta^2=\Var(\hat{o})/(M\delta^2)\leq \epsilon$, where $\delta$ bounds the additive error in $o_\text{avg}$ and $M$ is the number of identity copies of $\rho$ used. In other words, the number of experiments needed to achieved the desired tomography accuracy is given by 
\eq{\label{eq:Mbound}M\geq \Var(\hat{o})/\epsilon\delta^2.}
Therefore the problem boils down to analyzing the variance $\Var(\hat{o})$ of the single-shot random variable $\hat{o}=\Tr(O\hat{\rho})$. We will assume the physical observable $O$ to be traceless ($\Tr O=0$), since adding $O$ by $c\id$ only shifts $\hat{o}$ by a constant $c$ (given $\Tr(\hat{\rho})\equiv1$), which does not affect its variance. 

We can further bound the variance by
\eqs{\label{eq:var_definition}
\Var(\hat{o})=\dsE[\hat{o}^2]-\dsE[\hat{o}]^2\leq\dsE[\hat{o}^2],
}
By decomposing the observable $O=O_o+O_d$ to its off-diagonal part $O_o$ and traceless diagonal part $O_d$, we can evaluate $\dsE[\hat{o}^2]$, which takes the form of
\eqs{\label{eq:Vbound}\dsE[\hat{o}^2]=V[O^{\otimes2}]&=V[O_o^{\otimes2}]+V[O_d^{\otimes2}]+V[O_o\otimes O_d],}
where $V[O^{\otimes2}]$ is a linear function of the double-operator $O^{\otimes2}$, whose explicit form is given in \appref{appen:diagrammatic_variance}. More explicitly, we can express $V[O_o^{\otimes2}]=\Tr(O_o^2)F_o(t)$ and $V[O_d^{\otimes2}]=\Tr(O_d^2)F_d(t)$ with the dynamic form factors given by
\eqs{
&F_o(t)=f_1(t)+f_2(t)\dfrac{\Tr(O_o^2\rho)}{\Tr(O_o^2)}+f_3(t)\dfrac{\Tr(O_o^2\rho_d)}{\Tr(O_o^2)},\\
&F_d(t)=f_4(t)+f_5(t)\dfrac{\Tr(O_d^2\rho)}{\Tr(O_d^2)},
}
where the time-dependent functions $f_{1,2,3,4,5}$ are given in \eqref{eq:formfactors} of \appref{appen:diagrammatic_variance}. Combining the results in \eqnref{eq:Mbound}, \eqnref{eq:var_definition} and \eqnref{eq:Vbound}, we arrive at the following theorems regarding the efficiency of Hamiltonian-driven shadow tomography, which are central results of this work.

\begin{theorem}
Given an off-diagonal operator $O_o$, the Hamiltonian-driven shadow tomography with an evolution time $t$ uses $\scO(\frac{1}{\epsilon\delta^2}\Tr(O_o^2)F_o(t))$ independent copies of $\rho$ to estimate the expectation value $o_o$ of the observable $O_o$ to the precision that $\text{Pr}(|o_o-\Tr(O_o\rho)|\geq \delta)\leq \epsilon$.
\end{theorem}

\begin{theorem}
Given a traceless diagonal operator $O_d$, the Hamiltonian-driven shadow tomography with an evolution time $t$ uses $\scO(\frac{1}{\epsilon\delta^2}\Tr(O_d^2)F_d(t))$ independent copies of $\rho$ to estimate the expectation value $o_d$ of the observable $O_d$ to the precision that $\text{Pr}(|o_d-\Tr(O_d\rho)|\geq \delta)\leq \epsilon$.
\end{theorem}

In the long-time limit ($t\to \infty$),
\eq{F_o(\infty)=F_d(\infty)=1+2\frac{\Tr(O^2\rho)}{\Tr(O^2)}\leq 3.}
There is no difference between diagonal and off-diagonal observables in terms of tomography efficiency. The required number of samples (i.e.~copies of $\rho$) scales as $M\sim \Tr(O^2)/(\epsilon\delta^2)$ which agrees with the result for shadow tomography using Haar/Clifford random unitaries.\cite{Huang2020Predicting} For rank-1 observables, such as quantum fidelity, $\Tr(O^2)$ is independent of the system size $N$, then the shadow tomography only needs a constant number of experiments to achieve the desired accuracy.

We dubbed $F_o(t)$ and $F_d(t)$ as \emph{sample form factors} since they can be interpreted as the ratio of the required number of samples in the Hamiltonian-driven shadow tomography to that in the random-unitary-based shadow tomography (with 2-design unitary channels). A sample form factor less (or greater) than its long-time limit indicates the Hamiltonian-driven shadow tomography is more (or less) efficient than the random-unitary-based shadow tomography. With these understandings, we investigate the early time behavior of $F_o(t)$ and $F_d(t)$.

\begin{figure}[htb]
\centering
\includegraphics[width=0.8\linewidth]{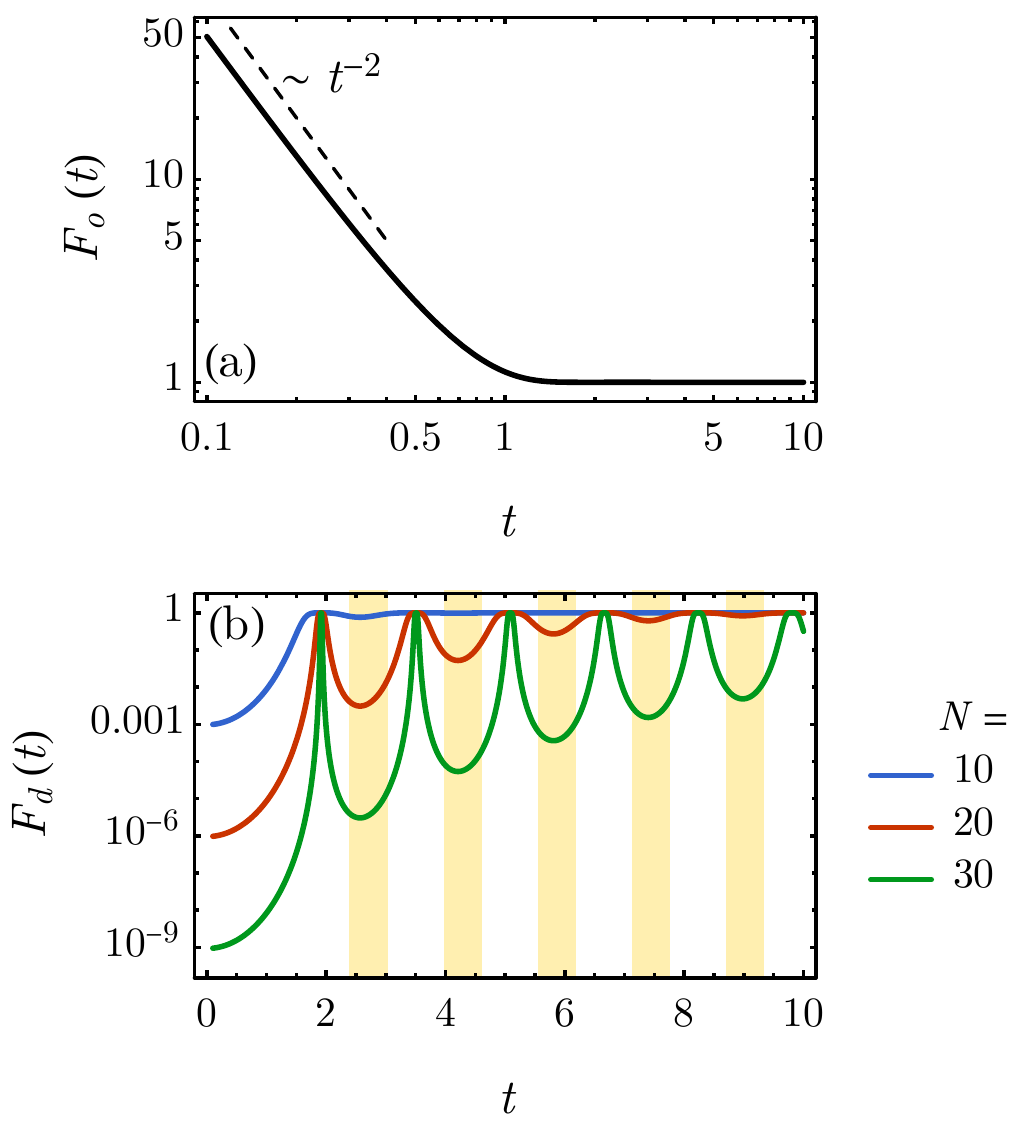}
\caption{Time-dependence of the sample form factor: (a) $F_o(t)$ for off-diagonal observables and (b) $F_d(t)$ for diagonal observables. The time scales in unit of the inverse energy scale of the chaotic Hamiltonian $H$.}
\label{fig:variance}
\end{figure}

At early time, the behavior of sample form factors can be rather complicated. However, we found that for Pauli observables (i.e.~$O$ is a Pauli operator), they take particularly simple forms (to the leading order in $D$)
\eqs{
F_o(t)=\dfrac{1}{1-r^{4}(t)},~F_d(t)=\dfrac{1}{1+Dr^{4}(t)}.
}
Recall that $r(t)=J_1(2t)/t$, the time-dependence of $F_o(t)$ and $F_d(t)$ for Pauli observables are plotted in \figref{fig:variance}. In the following, we will mainly focus on the Pauli observables. The general cases are discussed in Appendix \ref{appen:diagrammatic_variance} and \ref{appen:numerics}.

For off-diagonal observables, the sample form factor $F_o(t)$ diverges as $t\to0$, as it is impossible to infer the off-diagonal information from computational basis measurement in the absence of information scrambling. As time evolves, the off-diagonal information gets scrambled to the diagonal part, then $F_o(t)$ decays with $t$ as $t^{-2}$, as shown in \figref{fig:variance}(a). $F_o(t)$ quickly approaches $1$ after a characteristic time $t_o\sim 1$ set by the inverse energy scale of the Hamiltonian $H$, which was identified as the scrambling time of the system in \refcite{You2018Entanglement}. Thus in the Hamiltonian-driven shadow tomography, one just needs to wait for the scrambling time to achieve effectively the same efficiency as random-unitary-based shadow tomography in terms of off-diagonal observable. Moreover, the scrambling time $t_o$ is independent of $D$ (or the system size $N$) in our model, given the non-local nature of the GUE random Hamiltonian. 

For diagonal observables, the sample form factor $F_d(t)$ is of the order $F_d(t)\sim D^{-1}=2^{-N}$ at $t=0$, which is exponentially small in system size $N$. Without any unitary scrambling, the computational basis measurement is directly measuring the diagonal information of the density matrix, therefore it requires much fewer samples to infer diagonal observables as compared to that of the general-purpose random-unitary-based shadow tomography. As time evolves, the diagonal information is scrambled away, hence more samples are required to achieve the accuracy goal. So the tomography efficiency decreases with time for diagonal observables, in contrast to the increasing efficiency for off-diagonal observables.

Interestingly, $F_d(t)$ peaks to its maximal value periodically before it saturates to its long-time limit, as shown in \figref{fig:variance}(b). At these peaks, the diagonal information is maximally scrambled, therefore we name this phenomenon as \emph{scrambling beats}, which was first reported in \refcite{You2018Entanglement}. The peaks occur at times $t_k=x_k/2$, where $x_k$ is the $k$th zero of the Bessel function $J_1(x)$. Under coherent Hamiltonian evolution, the scrambled information can partially bounce back in a finite-size system, leading to the beat behavior of $F_d(t)$. But how long scrambling beats will last depends on the system size. The characteristic time for scrambling beats to die off is of the order $t_d\sim D^{1/6}=2^{N/6}$, when $Dr^{4}(t_d)\sim 1$. Before this time scale, there exist time windows between peaks, as yellow-shaded regions in \figref{fig:variance}(b), when the sample form factor maintains at a low level of $F_d(t)\sim D^{-1}$.

For a large enough system, the time scales $t_d\sim D^{1/6}$ and $t_o\sim 1$ are well-separated, which admits an intermediate time range $t_o\lesssim t \lesssim t_d$ where the Hamiltonian-driven shadow tomography can simultaneously achieve exponentially higher efficiency for diagonal observables and the same efficiency for off-diagonal observables, as compared to the random-unitary-base shadow tomography. Such behavior could potentially be advantageous when the diagonal observables are of more interest in certain tomography tasks.

\emph{Efficiency for Nonlinear Functions.}--- Our result can be generalized to analyze the tomography efficiency of predicting nonlinear functions of the density matrix $\rho$. For nonlinear function involving $k$ copies of $\rho$, which generally takes the form of $\Tr(O\rho^{\otimes k})$, the variance of the shadow estimation can be bounded by
\eqs{
&\Var(\Tr(O\hat{\rho}^{\otimes k}))\leq \dsE[\Tr(O\hat{\rho}^{\otimes k})^2]\\
&\leq \mathop{\sum}_{\vect{\alpha}}\Tr(O_{\alpha^L_1\alpha^L_2\cdots\alpha^L_k}O_{\alpha^R_1\alpha^R_2\cdots\alpha^R_k})\mathop{\prod}_{{\alpha}_i}F_{{\alpha}_i}(t),
}
where $\vect{\alpha}=(\alpha_1,\cdots,\alpha_k)$ has $k$-component, and each $\alpha_i=(\alpha^L_i,\alpha^R_i)\in\{(o,o),(d,d),(I,o),(o,I),(I,d),(d,I),(o,d),(d,o)\}$ is a pair of labels where $\alpha^{L/R}_i$ labels the $i$-th tensor leg of left/right $O$ operator. The summation is over all combinations of $\vect{\alpha}$. The details are discussed in Appendix \ref{appen:variance_nonlinear}. Plugging the variance into \eqnref{eq:Mbound}, the number of required experiments can be bounded as well for nonlinear functions. 

\emph{Summary and Discussions.}--- We propose to use Hamiltonian generated unitary evolution to scramble the quantum information for shadow tomography. We provide an unbiased estimator of the density matrix for all ranges of evolution time. We investigated the efficiency of the Hamiltonian-driven shadow tomography. In particular, for Pauli observables, we showed that it can be superior to the shadow tomography based on 2-design random unitaries within an intermediate time window. Although our analysis is based on the GUE random Hamiltonian, we expect that the result could be generalized to other types of quantum chaotic Hamiltonians\cite{Soonwon1,Soonwon2}. One interesting possibility is to consider Hamiltonians consist of random Pauli strings with random coefficients. In the strong disorder regime, such Hamiltonians can be asymptotically diagonalized by Clifford unitaries using the spectrum bifurcation renormalization group approach\cite{You2016Entanglement}, which enables efficient classical post-processing of the classical shadow data and makes the Hamiltonian-driven shadow tomography computationally feasible. Machine learning techniques may also find useful application in the classical post-processing phase to construct unbiased reconstruction channels based on data statistics, which helps to mitigate the influence of noises and imperfections of NISQ devices. Finally, we would like to mention that our current analysis is limited to non-local Hamiltonians. How to include locality into the discussion will be a challenging problem for future research.

\emph{Acknowledgements.}--- We thank the insightful discussion with Hsin-Yuan Huang, Junyu Liu, and Soonwon Choi. We also thank Richard Kueng for comments on the paper. HYH and YZY are supported by a startup fund from UC San Diego.


\bibliography{QST}

\newpage

\appendix
\onecolumngrid
\section{Diagrammatic approach towards quantum channel $\scM$\label{appen:diagrammatic_channel}}
In the shadow tomography of quantum states, for each experiment, the state $\rho$ is first evolved by a random unitary operator $U=e^{-i H t}=V\Lambda(t)V^{\dagger}$ generated from GUE random Hamiltonian $H$, then measured in the computational basis. The measurement will prepare a classical snapshot of quantum system $\hat{\sigma}=U^{\dagger}\ket{b}\bra{b}U$ with $b\in\{0,1\}^{N}$ labeling the measurement outcome. We can view the average mapping from the quantum state $\rho$ to its classical snapshots as a quantum channel $\scM$,
\eqs{
\mathcal{M}(\rho)&=\dsE[U^{\dagger}\ket{b}\bra{b}U]\\ &= \mathop{\dsE}_{V,\Lambda}\sum_{b\in\{0,1\}^N}V\Lambda(t)^\dagger V^\dagger\ket{b}\bra{b}V\Lambda(t) V^\dagger
\bra{b}V\Lambda(t) V^\dagger\rho V\Lambda(t)^\dagger V^\dagger\ket{b}\\
&=\mathop{\dsE}_{V,\Lambda}\dia{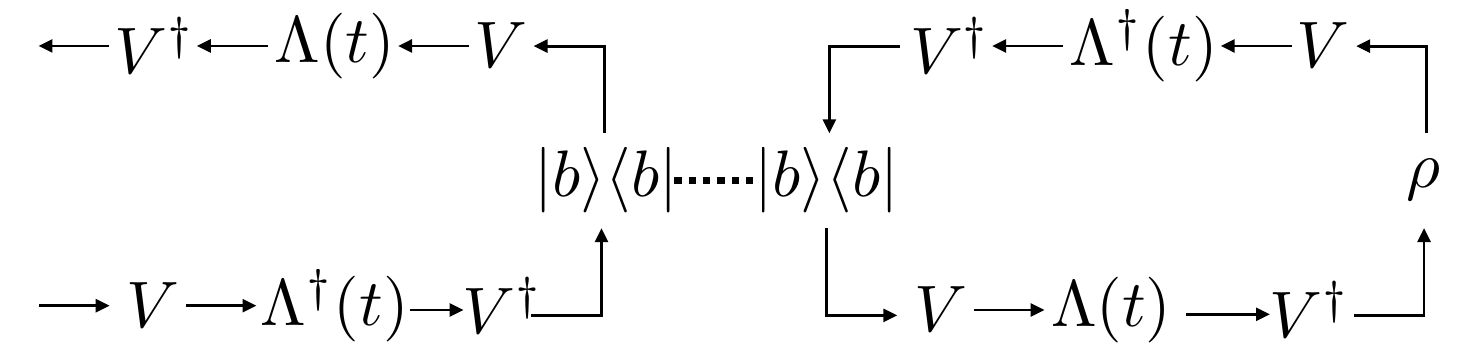}{55}{-27},}
in the last line we use the tensor network to represent the equation, the dashed line means summing over the bit-string $b$, and the arrow indicates the order that operators are multiplied together. As we can see this is a 4-fold twirl of Haar random matrix $V$, and $\Lambda(t)$ is a diagonal matrix with $\Lambda_{nn}(t)=e^{-i E_n t}$ with $E_n$ being eigen energies of $H$. It can be further simplified as
\eq{\scM(\rho)=\mathop{\dsE}_{\Lambda}\mathop{\sum}_{\sigma,\tau\in S_4}\text{Wg}[\sigma\tau^{-1} g_0]A[\sigma]B[\tau],}
where $\sigma,\tau$ are permutations from permutation group $S_4$, $\text{Wg}[g]$ is the Weingarten function\cite{Weingarten1978Asymptotic} of the permutation group element $g$, $g_0=(1,3)(2,4)$ is a fixed permutation to match the tensor network connection, and $A[\sigma]$, $B[\tau]$ are defined as:
\eqs{
A[\sigma]=\dia{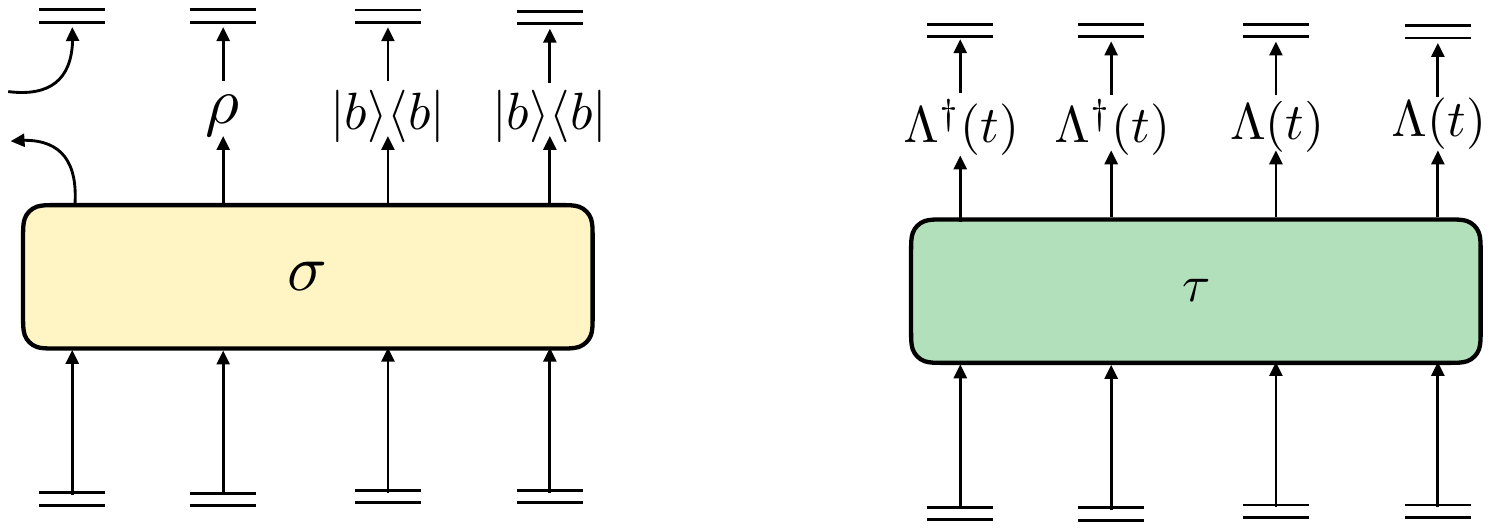}{90}{-45},~B[\tau]=\dia{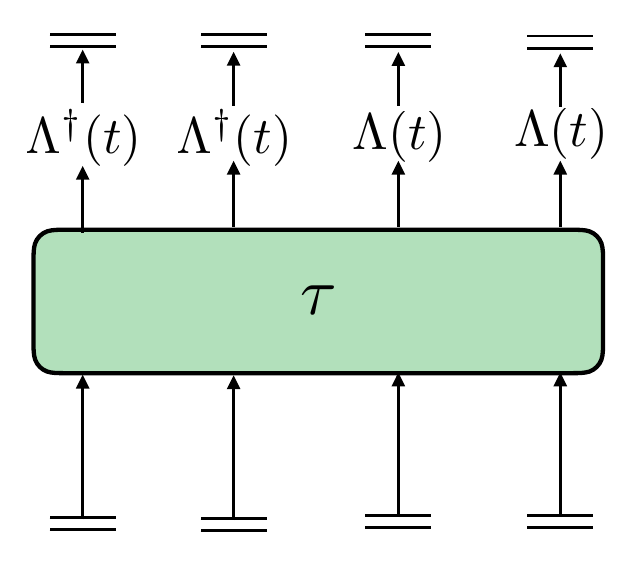}{90}{-45}.
}
In the above tensor diagram, the double line means the periodic boundary condition for top and bottom legs. After specifying the choice of permutation $\sigma$ and $\tau$, one can make a connection according to $\sigma$ and $\tau$ in the yellow and green block, and those tensor diagram can be evaluated.

The average over $\Lambda_{nn'}(t)=e^{-i E_n t}\delta_{n,n'}$ can be calculated using the joint probability distribution of eigen energies, and the spectral form factor of GUE matrices\cite{Cotler2017Chaos,You2018Entanglement}.
\eq{
P_{\text{GUE}}[E]\propto \mathop{\prod}_{m>m^{'}}\left(E_m-E_{m^{'}}\right)^{2}e^{-\frac{D}{2}\sum_m E_m^2}.
}
The summation of $\sigma,\tau\in S_4$ permutation group and averaging over eigen energies can be carries out and gives
\eqs{
\mathcal{M}(\rho)&=\mathcal{M}(\id/D+\rho_o+\rho_d)\\ 
 &=\dfrac{\mathbb{1}}{D}+\dfrac{\rho_o}{\alpha_D(t)}+\dfrac{\rho_d}{\beta_D(t)},
}
where $\id$ stands for the identity matrix, $\rho_o$ is the off-diagonal part of $\rho$, and $\rho_d$ is the traceless diagonal part of $\rho$. The coefficients $\alpha_D(t)$, and $\beta_D(t)$ are defined as
\eqs{
\alpha_D(t) &= \Big(\dfrac{1}{D+1}-\lambda_D(t)\Big)^{-1},\\
\beta_D(t)&=\Big(\dfrac{1}{D+1}+D\lambda_D(t)\Big)^{-1},\\
\lambda_D(t)&=\dfrac{(Dr^2(t)+r(2t))^2-4r^{2}(t)}{(D+3)(D^2-1)},}
and $r(t)=J_1(2t)/t$ with $J_1$ being the Bessel function of the first kind.

\section{Diagrammatic approach towards variance calculation\label{appen:diagrammatic_variance}}
\begin{figure}[H]
\centering
\includegraphics[width=\linewidth]{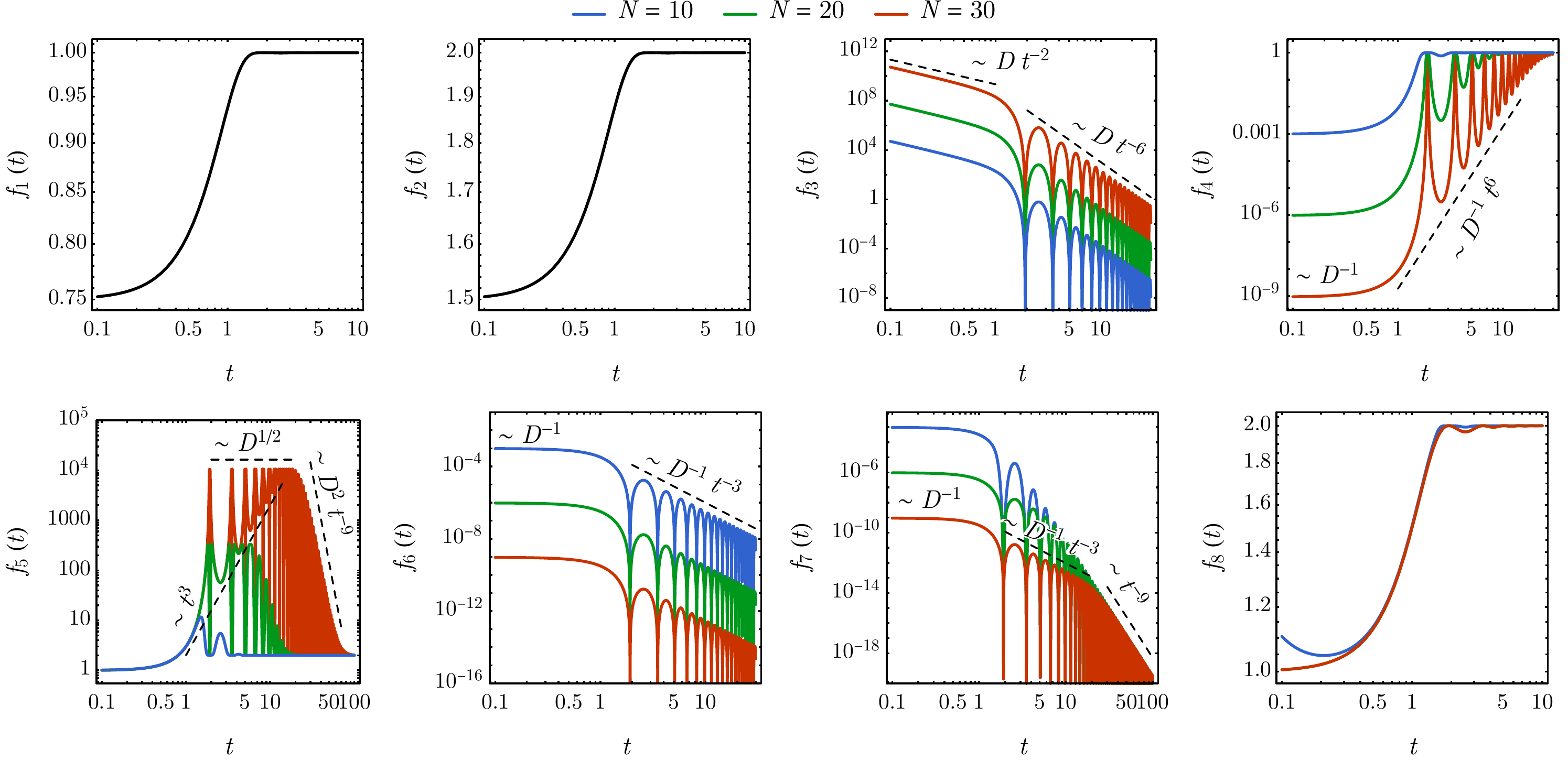}
\caption{Form factors $f_\alpha(t)$ and their asymptotic behaviors. Colored curves corresponds to different $D=2^N$ with $N=10,20,30$ (blue, green, red). For form factors independent of $D$, the curve is plotted in black.}
\label{fig:form factors}
\end{figure}

In the main text, we have shown that the efficiency of using shadow tomography to predict physical observables $o=F(O, \rho)$ is closely related to the variance of $\hat{o}=F(O,\hat{\rho})$, where $\hat{\rho}$ is the classical shadow, and $F$ is a function that depends on density matrix $\rho$, and observable $O$. In general, we have
\eq{\Var(\hat{o})\leq \dsE[\hat{o}^2],}
and we define a $V[\cdots]$ linear function of the double-operator $O^{\otimes 2}$ as
\eq{\dsE[\hat{o}^2]=\dsE[\Tr(O\hat{\rho})^2]=V[O^{\otimes 2}].}
Diagrammatically, the above equation can be expressed as 
\eq{
\dia{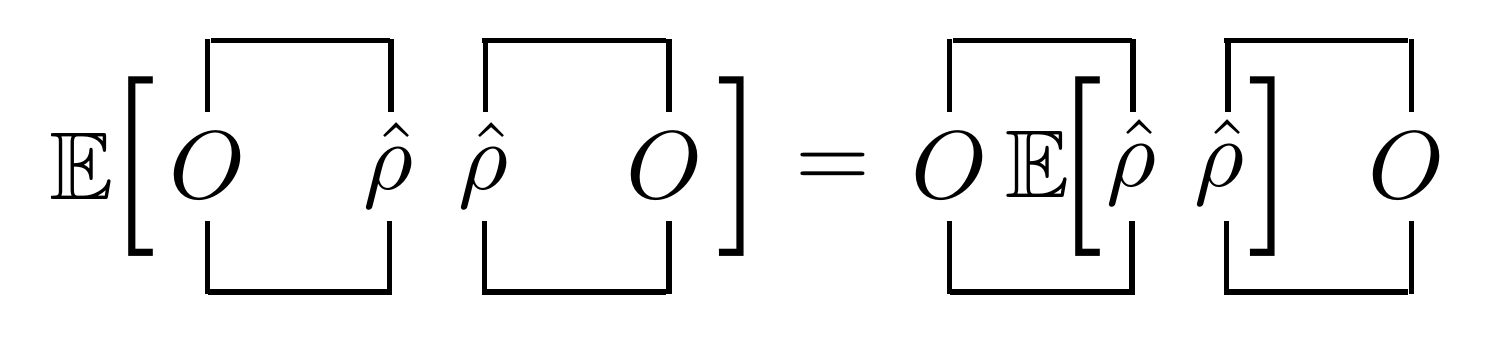}{35}{-5}.
}
The $V[\cdots]$ function that take $O^{\otimes 2}$ as input can be expressed diagrammatically as 
\eq{\dia{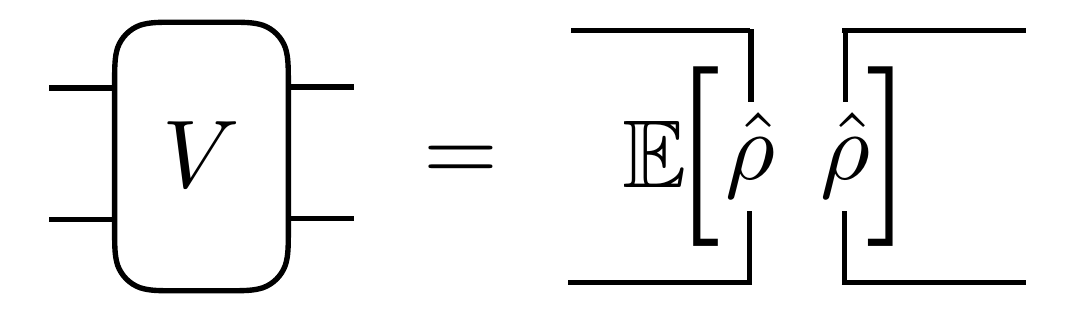}{35}{-5}.\label{appen:V_dia}}
The calculation of Eq.\ref{appen:V_dia} involves 6-design random unitaries and we showed that is involves 8 diagrams:
\eq{\dia{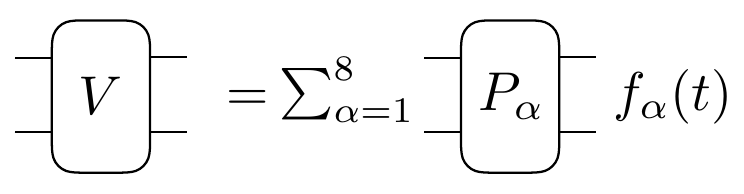}{35}{-5}.\label{eq:V}}
We name them P-diagrams, which are defined as:
\eqs{
&\dia{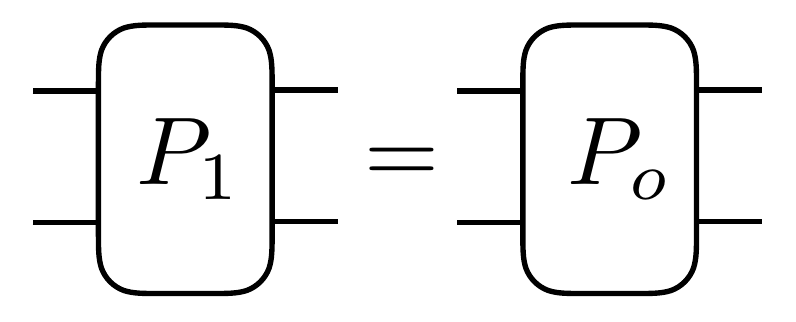}{35}{-5},\dia{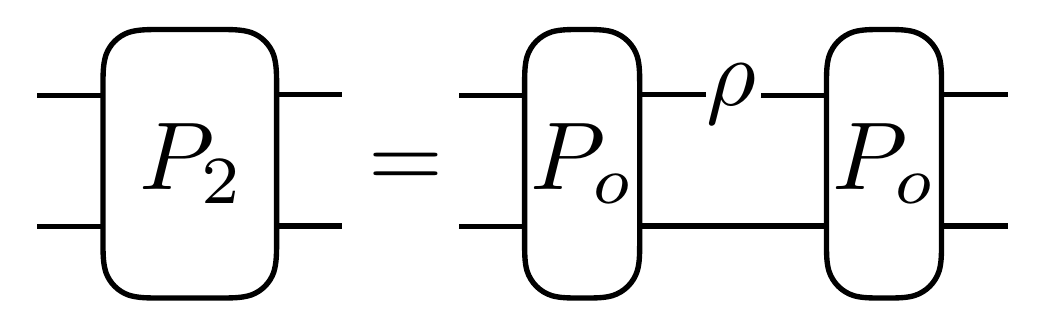}{35}{-5},\dia{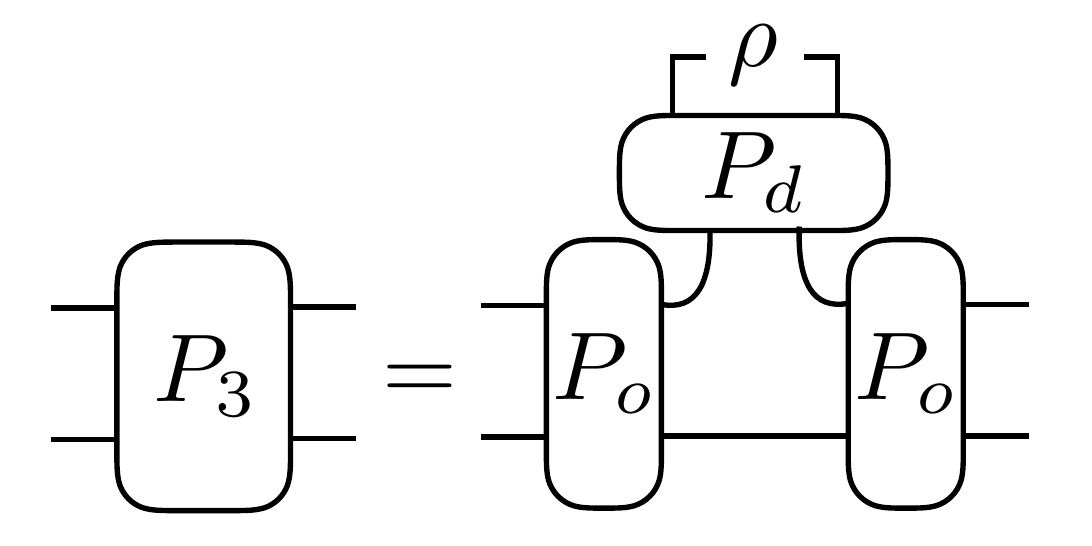}{60}{-5},\\
&\dia{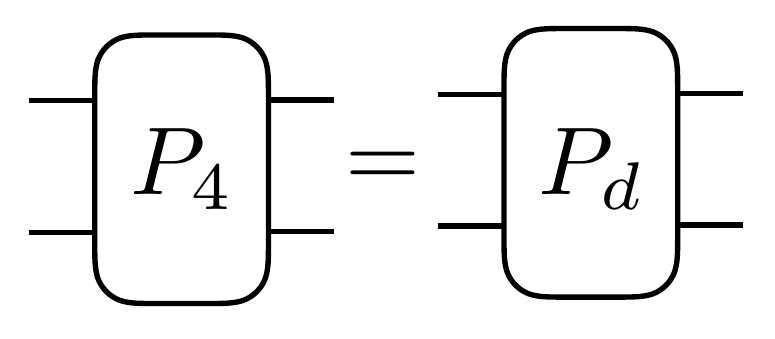}{35}{-5},\dia{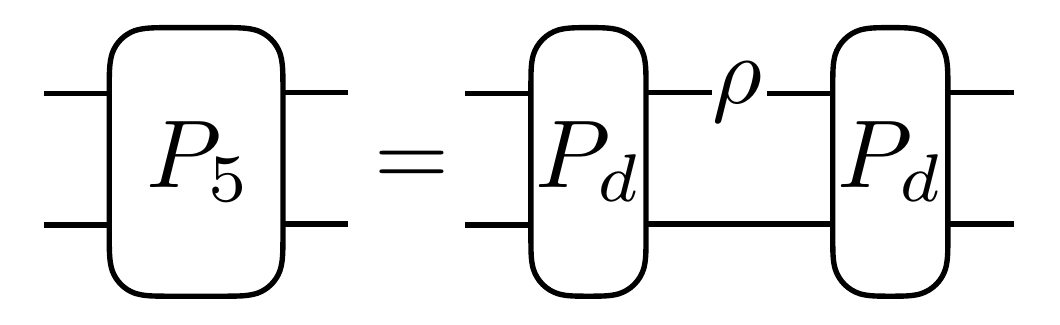}{35}{-5},\dia{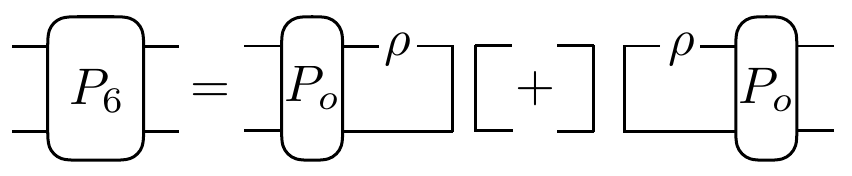}{38}{-5},\\
& \dia{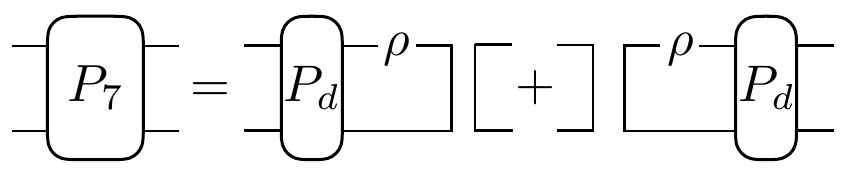}{38}{-5},\dia{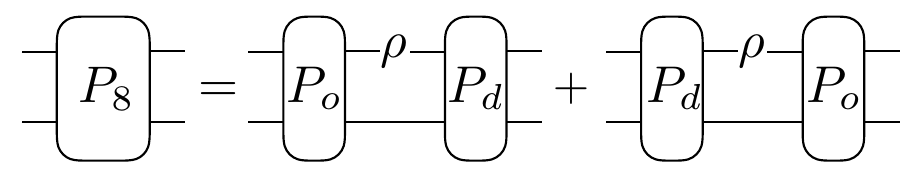}{38}{-5}\label{appen:V_dia2},
}
where $P_o$ is a projection operator that takes the off-diagonal part of an operator, $P_d$ is a projection operator that takes the traceless diagonal part of an operator. The \emph{form factors} $f_{\alpha}(t)$ are given by (with $r(t)=J_1(2t)/t$)
\eqs{\label{eq:formfactors}
&f_1(t)=\dfrac{1-3r^4(t)+2r^6(t)}{(1-r^4(t))^2}\\
&f_2(t)=2f_1(t)\\
&f_3(t)=\dfrac{2D(r^4(t)-r^6(t))}{(1-r^4(t))^2}\\
&f_4(t)=\dfrac{1}{1+Dr^4(t)}\\
&f_5(t)=\dfrac{2+D^2 r^6(t)+6Dr^{4}(t)}{(1+D r^{4}(t))^2}\\
&f_6(t)=\dfrac{r^2(t)-r^6(t)}{D(1-r^4(t))}\\
&f_7(t)=\dfrac{r^6(t)}{1+D r^{4}(t)}\\
&f_8(t)=\dfrac{2D(r^4(t)-r^6(t))+2}{(1+Dr^4(t))(1-r^4(t))}.
}
The shapes of these form factors are shown in \figref{fig:form factors}.

Collecting the P-diagrams,  $V[O^{\otimes2}]$ can be decomposed into five parts:
\eq{V[O^{\otimes2}]=\dia{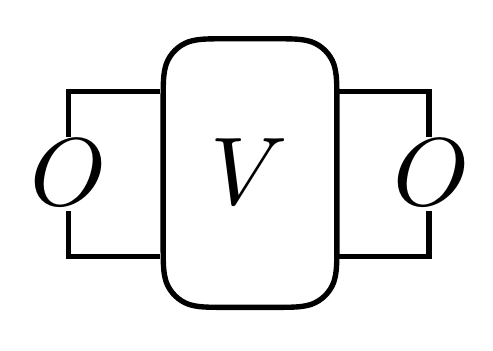}{35}{-15}=V[O_o^{\otimes2}]+V[O_d^{\otimes2}]+V[O_o\otimes O_d]+2V[\mathbb{1} \otimes O_o]+2V[\mathbb{1} \otimes O_d]}
where $V[O_o^{\otimes2}]$ involves diagram $P_1$, $P_2$, and $P_3$; $V[O_d^{\otimes2}]$ involves diagram $P_4$, and $P_5$; $V[\mathbb{1} \otimes O_o]$ involves $P_6$; $V[\mathbb{1} \otimes O_d]$ involves $P_7$;  $V[O_o\otimes O_d]$ involves $P_8$. Here $O_d$, $O_o$ stand for the traceless diagonal part and the off-diagonal part of the observable $O$. More explicitly, we can write down:
\eqs{
&V[O_o^{\otimes2}]=f_1(t)\Tr(O_o^2)+f_2(t)\Tr(O_o^2\rho)+f_3(t)\Tr(O_o^2\rho_d)\\
&V[O_d^{\otimes2}]=f_4(t)\Tr(O_d^2)+f_5(t)\Tr(O_d^2\rho)\\
&V[\mathbb{1} \otimes O_o]=f_6(t)\Tr(O_o \rho)\Tr(O)\\
&V[\mathbb{1} \otimes O_d]=f_7(t)\Tr(O_d \rho)\Tr(O)\\
&V[O_o\otimes O_d]=f_8(t)(\Tr(O_d O_o \rho)+\Tr(O_o O_d \rho)).
}
We analyze the asymptotic behavior of form factor $f_{\alpha}(t)$ in the following, and provide numerical evidence in the following appendix. Given $r(t)=J_1(2t)/t$ and the envelop behavior of the Bessel function $J_\nu(t)\sim\sqrt{2/(\pi t)}$, we can obtain the following asymptotic behavior
\eq{r(t)\overset{t\to 0}{=}1-\frac{1}{2}t^2,\quad r(t)\overset{t\to\infty}{\simeq}t^{-3/2}.}

\emph{Off-diagonal terms:} The off-diagonal terms involve $P_1$, $P_2$, and $P_3$, corresponding to the form factors $f_1$, $f_2$ and $f_3$. As shown in \figref{fig:form factors}, the form factors $f_1(t)$ and $f_2(t)$ are bounded between their $t=0$ and $t\to\infty$ values,
\eqs{
&\lim_{t\rightarrow 0}f_1(t)=0.75,~\lim_{t\rightarrow\infty}f_1(t)=1,\\
&\lim_{t\rightarrow 0}f_2(t)=1.5,~\lim_{t\rightarrow\infty}f_2(t)=2.
}
The early-time divergence behavior of off-diagonal terms comes from $f_3$. As we can see,
\eq{
f_3(t)\overset{t\to 0}{=}\frac{1}{2}Dt^{-2}.
}
So for short time, $f_3(t)$ diverges as $t^{-2}$ as shown in \figref{fig:form factors}. The long time behavior of $f_3(t)$ is
\eqs{f_{3}(t)\overset{t\rightarrow\infty}{\simeq} D t^{-6}.}

\emph{Diagonal traceless terms:} The diagonal traceless terms involve two P-diagrams, $P_4$ and $P_5$, corresponding to $f_4$ and $f_5$ form factors.

For $f_4(t)$, the form factor reaches local maximal $f_4(t_k)=1$ at $t_k=x_k/2$, where $x_k$ is the $k$th zero point of the Bessel function $J_1(x)$, see \figref{fig:form factors}. And in the window between those points, $f_4(t)\sim D^{-1}t^6$. We name this phenomenon as \emph{scrambling beats}. And for the long time,
\eqs{f_{4}(t)\overset{t\rightarrow \infty}{\simeq}1-D t^{-6}.}
Therefore the beats behavior of $f_4(t)$ will last for a characteristic time $T_4\sim D^{1/6}$.

For $f_5(t)$, in the long time, it becomes
\eqs{
f_5(t)\overset{t\to\infty}{\simeq}2+2D t^{-6}+D^2 t^{-9}.
}
We can define two characteristic times $T_5^{(1)}\sim D^{1/6}$ and $T_5^{(2)}\sim D^{2/9}$. In  \figref{fig:form factors}, we show the asymptotic behavior of $f_5(t)$. We can see the oscillation behavior separates to two stages. In the first stage, it will peak at a constant value that scales as $D^{1/2}$. This stage ends around $T_5^{(1)}\sim D^{1/6}$. Then, in the second stage, the peak of $f_5(t)$ will decay and eventually reaches its long-time value. Time $T_5^{(2)}\sim D^{2/9}$ characters the total time for $f_5(t)$ to reach its long-time value.

We now analyze how the peak value scales with $D$ in the first stage. We observe the peak happens in vicinity of $t_k=x_k/2$. First, we Talyor expand $r(t)$ around $t_k$ as
\eqs{
&r(t_k+\delta t)=a_k\delta t\\
&f_5(t_k+\delta t)=\dfrac{2+D^2 a_k^6 \delta t^6+6D a_k^4\delta t^4}{(1+Da_k^4 \delta t^4)^2}
}
By analyzing the behavior of $f_5(t_k+\delta t)$, we can find the maximal peak happens at $\delta t=\pm\frac{3^{1/4}}{a_k D^{1/4}}$ with the peak value $\text{max}(f_5)=\frac{5}{4}+\frac{3\sqrt{3}\sqrt{D}}{16}$. 

\section{Variance of linear functions\label{appen:variance_linear}}
With the diagrammatic tool developed in \appref{appen:diagrammatic_variance}, we can discuss the efficiency of shadow tomography in predicting linear and nonlinear functions. In this section, we are going to focus on the linear function prediction, which has the form $o=\Tr(O\rho)$, and in \appref{appen:numerics}, we will provide numerical evidence of variance behavior under various conditions. In addition, in \appref{appen:variance_nonlinear}, we will discuss the prediction of nonlinear function $o=\Tr(O\rho^{\otimes k})$, which involves $k$ copies of $\rho$. 

The linear function $o=\Tr(O\rho)$ can be estimated from classical shadows $\hat{\rho}$ via $o=\dsE[\Tr(O\hat{\rho})]=\dsE[\hat{o}]$, where $\hat{o}\equiv \Tr(O\hat{\rho})$ can be viewed as a random variable derived from the classical shadow. In the main text, we have shown the efficiency of shadow tomography is closely related the variance of estimation $\Var(\hat{o})$, which is upper bounded by 
\eq{\Var(\hat{o})\leq \dsE[\hat{o}^2].}
Diagrammatically, it can be expressed as
\eq{\dia{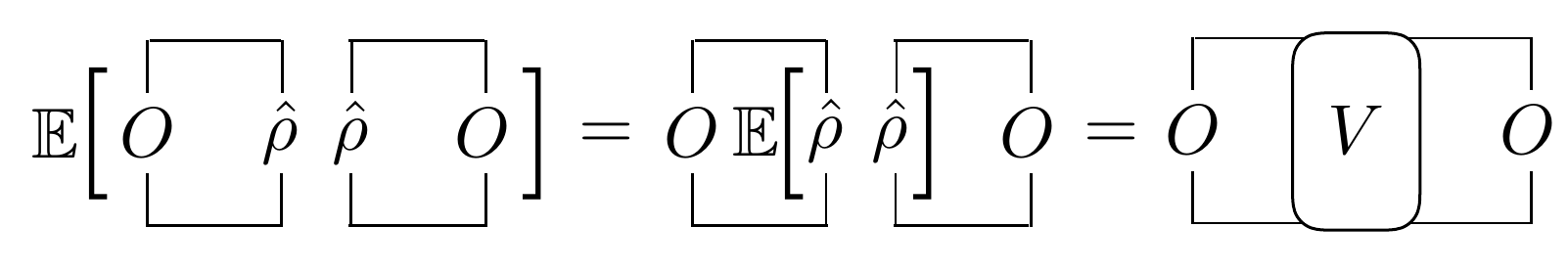}{35}{-5},}
where $V$ operator is defined in Eq.\ref{eq:V}. Furthermore, since $\Tr(\hat{\rho})=1$, adding O by $c\mathbb{1}$ only shifts $\hat{o}$ by a constant c, and it will not affect variance. So for linear function, we can assume the observable $O$ to be traceless, i.e. $\Tr(O)=0$. Then
\eqs{
\Var(\hat{o})\leq \dsE[\hat{o}^2]=V[O^{\otimes 2}]=V[O_o^{\otimes 2}]+V[O_d^{\otimes 2}]+V[O_o\otimes O_d].\label{eq:appen_linear_function}
}
\subsection{Off-diagonal Pauli observables}
We define the off-diagonal dynamical form factor $F_o(t)$ as
\eqs{
F_o(t)=\dfrac{V[O_o^{\otimes 2}]}{\Tr(O_o^2)}.
}
If the observables are off-diagonal Pauli operator $O_o$, then $F_o(t)$ can be simplified as
\eqs{
F_o(t)=\dfrac{f_1(t)\Tr(O_o^2)+f_2(t)\Tr(O_o^2\rho)+f_3(t)\Tr(O_o^2\rho)}{\Tr(O_o^2)}=\dfrac{f_1(t)D+f_2(t)+f_3(t)}{D}\simeq\dfrac{1}{1-r^{4}(t)}.
}
In the last step, we only keep the leading $D$ terms.
\subsection{Diagonal Pauli observables}
We define the diagonal dynamical form factor $F_d(t)$ as
\eqs{
F_d(t)=\dfrac{V[O_d^{\otimes 2}]}{\Tr(O_d^2)}.
}
If the observables are diagonal Pauli operator $O_d$, then $F_d(t)$ can be simplified as
\eqs{
F_d(t)=\dfrac{f_4(t)\Tr(O_d^2)+f_5(t)\Tr(O_d^2\rho)}{\Tr(O_d^2)}=\dfrac{f_4(t)D+f_5(t)}{D}\simeq f_4(t)=\dfrac{1}{1+D r^{4}(t)}.
}
In the second last step, we use the fact that $\max(f_5(t)/D)\sim D^{-1/2}$. For large $D$, this contribution can be ignored.

\section{Numerical results and case studies\label{appen:numerics}}
\subsection{Numerical studies of reconstruction channel}
\begin{figure}[H]
    \centering
    \includegraphics[width = 0.6\linewidth]{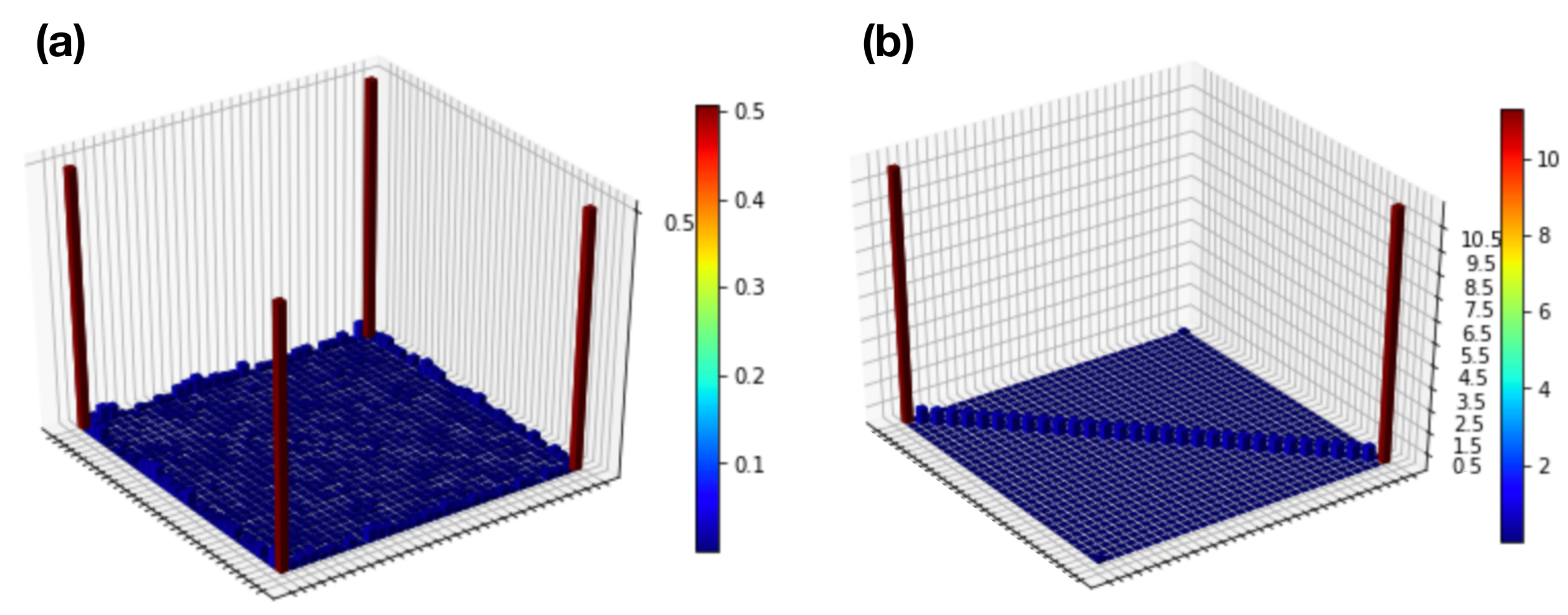}
    \caption{Early time reconstruction of GHZ state. Measurements are taken at $T=0.4$. (a) shows the unbiased reconstruction of GHZ density matrix with 100,000 classical snapshots, while (b) shows the reconstruction channel using unitary 2-design is highly biased for early time reconstruction.}
    \label{fig:reconstruction}
\end{figure}

In the main text, we derived the unbiased reconstruction channel $\scM^{-1}(X)$ for whole range of dynamical time. We demonstrated this using 5 qubits GHZ states, which GUE random Hamiltonians. At early time $T=0.4$, we collected $10^5$ classical snapshots. In \figref{fig:reconstruction}(a), we showed our reconstruction channel give unbiased density matrix of 5 qubits GHZ states, while the reconstruction channel of unitary 2-design is high biased for early time reconstruction. Especially, the off-diagonal information in density matrix is missing. In practice, we need to measure the state and collect classical snapshots at short-time or intermediate-time scale. Firstly, we want to shrink the evolution time to reduce the total time of experiments. Secondly, long-time evolution posts difficulties in classical post-processing. Because simulate chaotic Hamiltonian dynamics is hard and inaccurate for long time. These reasons justify the need for an unbiased reconstruction channel for whole time range.

\subsection{Numerical case studies of variance estimations}
In the main text, \appref{appen:diagrammatic_variance} and \appref{appen:variance_linear}, we derived and analyzed the upper bound of the variance of $\hat{o}=\Tr(O \hat{\rho})$. We numerically studied variance for several cases. We found our analytical calculation agrees well with numerical simulations, and the upper bound is tight for those cases.

\textbf{Case 1:} The observable only contains off-diagonal term $O = O_o$. 

\begin{figure}[H]
    \centering
    \includegraphics[width=0.8\linewidth]{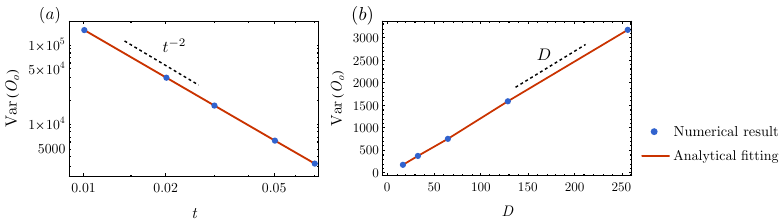}
    \caption{Numerical tests on the scaling of variance for off-diagonal operators. (a) shows the $t^{-2}$ scaling using 5 qubits GHZ states; (b) shows $D$ scaling at fixed evolution time $T=0.1$.}
    \label{fig:off_scaling}
\end{figure}

From the previous discussion, we know 
\eqs{
\Var(\hat{o})\leq\dsE[\hat{o}^2]=f_1(t)\Tr(O_o^2)+f_2(t)\Tr(O_o^2\rho)+f_3(t)\Tr(O_o^2\rho_d)
}
The $f_1(t)$ and $f_2(t)$ terms are regular and bounded, while $f_3(t)$ will give $t^{-2}$ divergence as
\eqs{
\lim_{t\rightarrow 0}f_3(t)=\dfrac{D}{2t^2}.
}
So in early time, $\dsE[\hat{o}]$ will scale with $D$, and $t^{-2}$. To test these phenomenon, we prepare the system in the GHZ state, and $O_o=0.5\ket{\uparrow\cdots\uparrow}\bra{\downarrow\cdots\downarrow}+0.5\ket{\downarrow\cdots\downarrow}\bra{\uparrow\cdots\uparrow}$ to be the off-diagonal fidelity. We collected $10^4$ classical snapshots, and numerically estimate $\Var(O_o)$ under various conditions. In \figref{fig:off_scaling}(a), we found the $t^{-2}$ behavior agrees well with the numerical simulation. And in \figref{fig:off_scaling}(b), we also numerically confirmed the variance scales with Hilbert space dimension $D$.

\textbf{Case 2:} The observable only contains diagonal term $O = O_d$, and $\Tr(O_d^2\rho)\simeq \Tr(O_d^2)/D$.

\begin{figure}[H]
    \centering
    \includegraphics[width=0.8\linewidth]{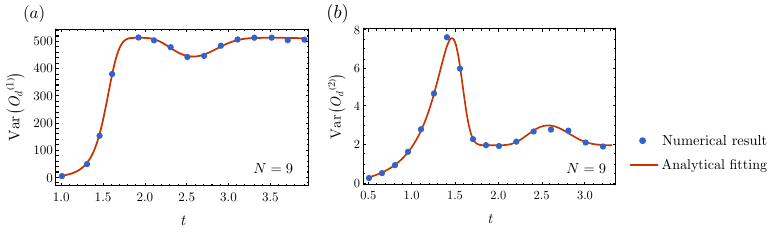}
    \caption{Caption}
    \label{fig:dia_scaling}
\end{figure}

For diagonal operator $O_d$, the variance of estimation using shadow tomography is upper bounded by
\eqs{
\Var(\hat{o}_d)\leq \dsE[\hat{o}_d^2]=f_4(t)\Tr(O_d^2)+f_5(t)\Tr(O_d^2\rho)=(f_4(t)+f_5(t)/D)\Tr(O_d^2).
}
The behavior of the variance will dominated by $f_4(t)$. To test this, we prepared 9 qubits GHZ state as initial state, and we measure the diagonal operator $O_d=Z_1I_2\cdots I_9$. In \figref{fig:dia_scaling}(a), the blue dots shows the numerical estimation of the variance of $\hat{o}_d$. Indeed, we see numerical results match $f_4(t)$ behavior. And to further quantify it, we use ansatz $c_1f_4(t)+c_2f_5(t)+c_3$ to fit the numerical results, where $c_{1,2,3}$ are fitting parameters. By minimizing the mean-square error, we got the best fitting results with $c_1=514.5$, $c_2=0.93$, and $c_3=-1.742$. We see the value of $c_1$ matches its theoretical value $c_1=\Tr(O_d^2)=512$, and $c_2$ matches $c_2=\Tr(O_d^2\rho)=1$ really well. And the red curve in \figref{fig:dia_scaling}(a) shows the fitting result.

\textbf{Case 3:} The observable only contains diagonal term $O = O_d$, and $\Tr(O_d^2\rho)\simeq \Tr(O_d^2)$.

If this is the case, then the variance behavior will be dominated by $f_5(t)$. To test this idea, we prepare the state in $\ket{\psi}=\ket{\downarrow\cdots\downarrow}$ with 9 qubits, and choose the diagonal operator as $O_d=\ket{\downarrow\cdots\downarrow}\bra{\downarrow\cdots\downarrow}$. In \figref{fig:dia_scaling}(b), we see the variance behavior is indeed resembles $f_5(t)$. We also use ansatz $c_1f_4(t)+c_2f_5(t)+c_3$ to fit the numerical results, where $c_{1,2,3}$ are fitting parameters. By minimizing the mean-square error, we got the best fitting results with $c_1=1.02$, $c_2=0.94$, and $c_3=-1.02$. Again, the fitting result $c_1$ matches its theoretical result $c_1=\Tr(O_d^2)=1$, and $c_2$ matches $c_2=\Tr(O_d^2\rho)=1$ very well. From these tests, we can see our analytical upper bound estimation of variance matches the numerical results under various conditions, and the upper bound is tight in the numerical tests.

\section{Variance of nonlinear functions\label{appen:variance_nonlinear}}
For nonlinear function involving $k$ copies of $\rho$, such as $o_k=\Tr(O\rho^{\otimes k})$, we can estimate it using $\hat{o}_k=\Tr(O\hat{\rho}_1\otimes\cdots \otimes\hat{\rho}_k)$, where $\hat{\rho}_i$ are independent samples of classical shadows. The efficiency is related to $\Var(\hat{o}_k)\leq \dsE[\hat{o}_k^2]$. Since $\hat{\rho}_i$ are independent random variables, we can have
\eq{
\dsE[\hat{o}_k^2]=\dia{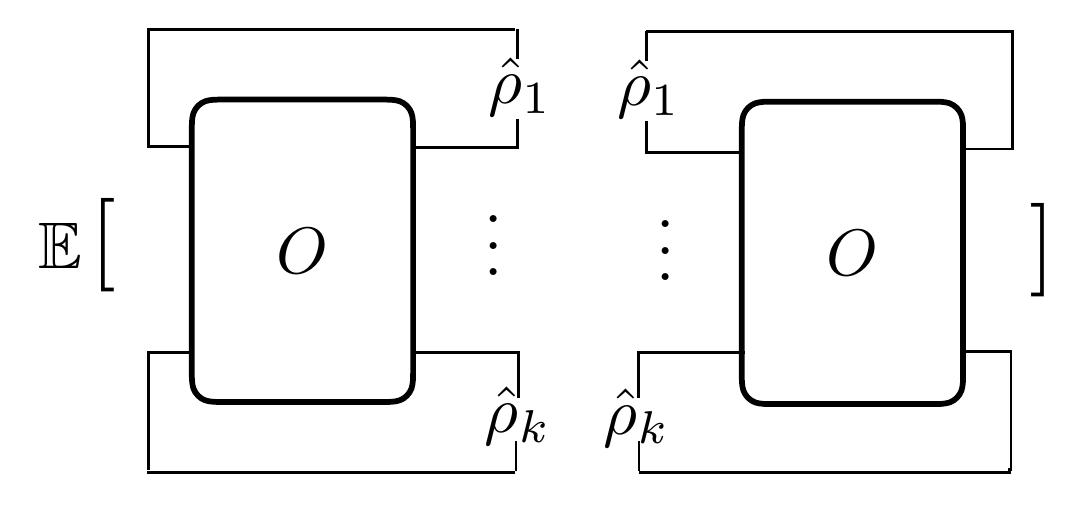}{85}{-40}=\dia{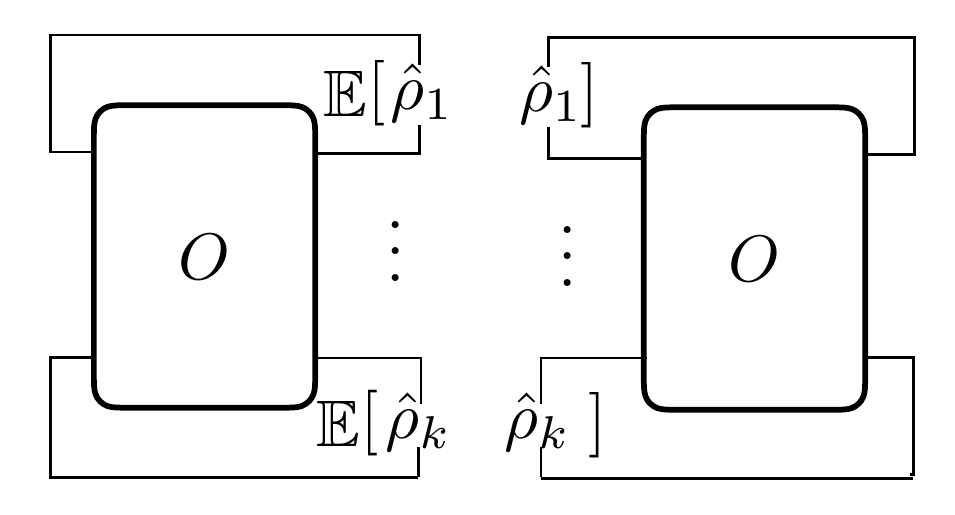}{80}{-38}.
}
Using Eq.\ref{appen:V_dia}, we can simplify the above equation as 
\eq{\dsE[\hat{o}_k^2]=\dia{M2}{80}{-38}=\dia{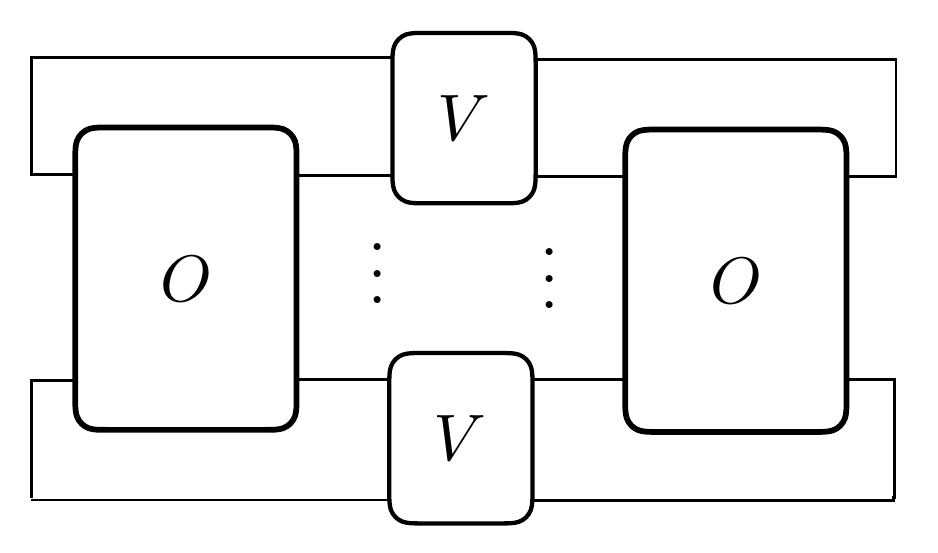}{80}{-38},\label{appen:nonlinear_var}}
where each $V[\cdots]$ involves 7 P-diagrams as defined in Eq.\ref{appen:V_dia2}. From Eq.\ref{appen:nonlinear_var}, we can in principle calculate the upper bound for variance of nonlinear functions involving any order $k$ of $\rho$:

\eqs{
\Var(\Tr(O\hat{\rho}^{\otimes k}))\leq \dsE[\Tr(O\hat{\rho}^{\otimes k})^2]= \sum_{\vect{\alpha}\in[1,8]^{k}}\prod_{\alpha_i}f_{\alpha_i}(t)P_{\alpha_i},\label{eq:p_diagram_exp}
}
where $\vect{\alpha}=(\alpha_1,\cdots, \alpha_k)\in[1,7]^{k}$ labels the P-diagram attached to each tensor leg. For convenience, we rename
\eqs{
&f_1(t)+f_2(t)+f_3(t)=F_{(o,o)}(t)\\
&f_4(t)+f_5(t)=F_{(d,d)}(t)\\
&f_6(t)=F_{(I,o)}(t)=F_{(o,I)}(t)\\
&f_7(t)=F_{(I,d)}(t)=F_{(d,I)}(t)\\
&f_8(t)=F_{(o,d)}(t)=F_{(d,o)}(t).
}
With those names, we can formally rewrite Eq.\ref{eq:p_diagram_exp} as
\eqs{
\Var(\Tr(O\hat{\rho}^{\otimes k}))\leq \dsE[\Tr(O\hat{\rho}^{\otimes k})^2]\leq\mathop{\sum}_{\vect{\alpha}=(\vec{\alpha}_1,\cdots,\vec{\alpha}_k)}\Tr(O_{\alpha^L_1\alpha^L_2\cdots\alpha^L_k}O_{\alpha^R_1\alpha^R_2\cdots\alpha^R_k})\mathop{\prod}_{\vec{\alpha}_i}F_{\vec{\alpha}_i}(t),
}
where $\vect{\alpha}=(\vec{\alpha}_1,\cdots,\vec{\alpha}_k)$ has $k$-component, and each $\vec{\alpha}_i=(\alpha^L_i,\alpha^R_i)$ is a two component vector where $\alpha^{L/R}_i$ labels the $i$-th tensor leg of left/right $O$ operator. Each $\vec{\alpha}_i=(\alpha^L_i,\alpha^R_i)\in\{(o,o),(d,d),(I,o),(o,I),(I,d),(d,I),(o,d),(d,o)\}$, and the summation is over all combinations of $\vect{\alpha}$. In the last step, we use the fact, $\Tr(O\rho)\leq\Tr(O)$.


\end{document}